\documentclass[11pt,superscriptaddress]{article}
\usepackage{jheppub}
\usepackage{epsfig}
\usepackage{amsmath,amssymb}
\usepackage{amsfonts}
\usepackage{mathrsfs}
\usepackage{units}
\usepackage{multirow}
\usepackage{relsize}
\RequirePackage{xspace}
\usepackage{graphicx}
\usepackage{slashed}
\usepackage{bm}

\title{Search for $C=+$ charmonium and $XYZ $  states in
$e^+e^-\to \gamma+~H$ at BESIII}

\author[a]{ Yi-Jie Li }
\author[a]{, Guang-Zhi Xu }
\author[b]{, Kui-Yong Liu }
\author[a,c]{, Yu-Jie Zhang}

\affiliation[a]{Key Laboratory of Micro-nano
 Measurement-Manipulation and Physics (Ministry of Education) and School of Physics,
  Beihang University, Beijing 100191, China}

\affiliation[b]{Department of Physics, Liaoning University, Shenyang 110036
, China}

\affiliation[c]{International Research Center for Nuclei and Particles in the Cosmos, Beihang University, Beijing 100191, China}

\emailAdd{ yijiegood@gmail.com}

\emailAdd{ still200@gmail.com}

\emailAdd{liukuiyong@lnu.edu.cn}

\emailAdd{nophy0@gmail.com}

\abstract{
Within the framework of nonrelativistic quantum chromodynamics,
we study the production of $C=+$ charmonium states $H$ in $e^+e^-\to
\gamma~+~H$ at BESIII with $H=\eta_c(nS)$ (n=1, 2, 3, and 4),
$\chi_{cJ}(nP)$ (n=1, 2, and 3), and $^1D_2(nD)$ (n=1 and 2).
The radiative and relativistic corrections are calculated
to next-to-leading order for $S$ and $P$ wave states.  We then argue that the
search for $C=+$ $XYZ$ states such as $X(3872)$, $X(3940)$,  $X(4160)$, and $X(4350)$ in $e^+e^-\to
\gamma~+~H$ at BESIII may help clarify the nature of
these states. BESIII can search $XYZ$ states through two body process $e^+e^-\to \gamma H$, where $H$ decay to $J/\psi \pi^+\pi^-$, $J/\psi \phi$, or $D \bar D$. This result may be useful in identifying the nature of $C=+$ $XYZ$ states.
For completeness, the production of  $C=+$ charmonium
in $e^+e^-\to \gamma +~H$ at B factories is also discussed.
}

\begin{document}
\maketitle
\flushbottom





\section{Introduction}
During the last 10 years, many heavy quarkonium or heavy quarkonium-like $XYZ$
states had been discovered (more details can be found in Ref.\cite{Brambilla:2010cs} and related papers).
The $X(3872)$ state is the first and the most famous state among them.
It was discovered by the Belle collaboration\cite{Choi:2003ue}, and
confirmed by the CDF \cite{Acosta:2003zx}, D0\cite{Abazov:2004kp}, BaBar\cite{Aubert:2004ns},
LHCb\cite{Aaij:2011sn},  and CMS\cite{Chatrchyan:2013cld} collaborations.
One of the most conspicuous  properties of  $X(3872)$ is its mass, which is  close to
the $D^0\bar{D}^{\star0}$ threshold within $1$~MeV; hence, $X(3872)$ is suggested to be a
$D^0\bar{D}^{\star0}$ molecule
\cite{Braaten:2003he,
Close:2003sg,Wong:2003xk,Voloshin:2003nt}. The contribution of the charged component $D^+D^{\star-}$ is also considered in Ref.\cite{Gamermann:2009uq,Aceti:2012cb}.
The molecule model may be puzzled to explain the production
cross-sections of $X(3872)$ in hadron colliders ( which may be large in some phenomenological models\cite{Artoisenet:2010uu} ) \cite{Suzuki:2005ha}.
The quantum numbers of $X(3872)$ have been determined
to be $J^{PC}=1^{++}$ by LHCb collaboration~\cite{Aaij:2013zoa}.
The $J^{PC}$ of $X(3872)$ is the same as $\chi_{c1}(nP)$.  On the contrary, the mass $3.872$~GeV seems too low for a
$\chi_{c1}(2P)$ state.
The coupled-channel and screening effects  may draw its mass down to $3.87$~GeV
\cite{Li:2009zu}. However, next-to-leading order (NLO) prediction of $X(3872)$ production in hadron colliders within
nonrelativistic quantum chromodynamics (NRQCD) disfavors the interpretation of $X(3872)$ as pure
$\chi_{c1}(2P)$ \cite{Butenschoen:2013pxa}.
The possibility that $X(3872)$ might be a mixture state with the $\chi_{c1}(2P)$
and the $D^0\bar{D}^{\star0}$ components was proposed in Ref.\cite{Meng:2005er}.
The prompt $X(3872)$ hadroproduction is studied at NLO in $\alpha_s$\cite{Meng:2013gga}
and the result is consistent with the CMS \cite{Chatrchyan:2013cld} and  the CDF data\cite{Acosta:2003zx}.
This idea is also favored the data of some other measurements and predictions
\cite{Suzuki:2005ha,Meng:2007cx,Li:2009ad,Li:2009zu
}.

Besides $X(3872)$, other $C=+$~ $XYZ$ states are listed in Table  \ref{tab:c+xyz}.
These states are particularly interesting
and the interpretations for their nature are still inconclusive\cite{Molina:2009ct}.
$X(3915)$ ($X(3945)$ or $Y(3940)$) and $Z(3930)$ are assigned as the
$\chi_{c0}(2P)$ and $\chi_{c2}(2P)$ states by the Particle Data Group\cite{Beringer:1900zz}. However this identification may be called into question\cite{Guo:2012tv}.
The experimental results for these $C=+$ states have induced
renewed theoretical interest in understanding the nature of charmonium-like
states. The
double charmonium production in $e^+e^-$ annihilation at
B factories\cite{Abe:2002rb,Aubert:2005tj} turned out to be a possible way
to identify the $C=+$ charmonium or charmonium-like states, recoiling
against the easily reconstructed $1^{--}$ charmonium $J/\psi$ and
$\psi(2S)$. In addition to  $\eta_c, \eta_c(2S)$, $\chi_{c0}$,
 $X(3940)$ (decaying into $D\bar {D^*}$), and $X(4160)$ (decaying
into $D^*\bar {D^*}$) have also been observed in double charmonium
production at B factories. However,  $\chi_{c1}$ and $\chi_{c2}$ states
are missing in production associated with $J/\psi$ at B factories.
Identifying the $C=+$ charmonium
states $H$ in the $e^+e^-\to
\gamma ^*\to \gamma+H$  process  at B factories is also proposed\cite{Chung:2008km,Braguta:2010mf}.
The quantum chromodynamics  (QCD) corrections of $e^+e^-\to
\gamma ^*\to \gamma+H$ at B factories are calculated in Ref.\cite{Li:2009ki,
Sang:2009jc}. The relativistic correction of  $e^+e^-\to
\gamma ^*\to \gamma+\eta_c$ is also included in Ref.\cite{Sang:2009jc}.
Indirect measurement of quarkonium in the two-photon process is also proposed\cite{Sang:2012cp}.

\begin{table*}[t]\label{tab:c+xyz}
\caption{$C=+$ $XYZ$ states.
$X(3915)$, $X(3945)$, and $Y(3940)$ is considered as $\chi_{c0}(2P)$
 for compatible properties.
$Z(3930)$ is considreed as $\chi_{c2}(2P)$\cite{Eidelman:2012vu,Brambilla:2010cs}.
 }
\setlength{\tabcolsep}{0.21pc}
\begin{center}
\begin{tabular}{lcclc}
\hline\hline
\rule[10pt]{-1mm}{0mm}
 State & $m (\Gamma)$~in MeV & $J^{PC}$ & \ \ \ \ Production~(Decay) &
     Ref  \\
\hline
\rule[10pt]{-1mm}{0mm}
$X(3872)$& 3871.68$\pm$0.17 ( $<1.2$) &
    $1^{++}$
    & $B\to K\, (\pi^+\pi^-J/\psi)$ &
    \cite{Choi:2003ue}  \\
&  & & $p\bar p\to (\pi^+\pi^- J/\psi)+ ...$ &
    \cite{Acosta:2003zx,Abulencia:2006ma}  \\
&  &   & $B\to K\, (\omega J/\psi)$ &
    \cite{Abe:2005ix,delAmoSanchez:2010jr} \\
&  & & $B\to K\, (D^0 \bar D^{*} )$ &
    \cite{Gokhroo:2006bt,Aubert:2007rva}  \\
&  & & $B\to K\, (\gamma J/\psi, \gamma \psi(2S))$ &
    \cite{Aubert:2006aj}\\
&  & &  $pp\to (\pi^+\pi^- J/\psi)+ ...$   &
    \cite{Aaij:2011sn,Chatrchyan:2013cld,Aaij:2013zoa}~   \\
$X(3915)$ & $3917.5\pm2.7$ ($27\pm 10$ )& $0^{++}$ &
    $B\to K\, (\omega    J/\psi   )$ &
    ~\cite{Abe:2004zs,Aubert:2007vj} \\
          & & & $e^+e^-\to e^+e^-\, (\omega   J/\psi   )$ &
    \cite{delAmoSanchez:2010jr,Lees:2012xs} \\
$X(3940)$ & $3942^{+9}_{-8}$ ( $37^{+27}_{-17}$ )& $J^{P+}$ &
     $e^+e^-\to J/\psi\,(  D   \bar D^*  )$ &
     \cite{Abe:2007sya} \\
$Y(4140)$ & $4143.0\pm3.1$  ( $12^{+9}_{-\ 6}$) & $J^{P+}$ &
     $B\to K\, (\phi J/\psi)$ &
     ~\cite{Aaltonen:2011at}~ \\
$X(4160)$ & $4156^{+29}_{-25} $ ( $139^{+110}_{-60}$) & $J^{P+}$ &
     $e^+e^-\to   J/\psi   \,(  D^{*+} \bar D^{*-}  )$ &
     ~\cite{Abe:2007sya}~ \\
$Y(4274)$ & $4274.4^{+8.4}_{-6.7}$ ( $32^{+22}_{-15}$) & $J^{P+}$ &
     $B\to K\, (\phi J/\psi)$ &
     ~\cite{Aaltonen:2011at}~   \\
$X(4350)$ & $4350.6^{+4.6}_{-5.1}$ ( $13.3^{+18.4}_{-10.0}$ )& 0/2$^{++}$ &
     $e^+e^-\to e^+e^- \,(\phi   J/\psi   )$ &
     ~\cite{Shen:2009vs}\\
\hline\hline
\end{tabular}
\end{center}
\end{table*}

Recently, BesIII reports the cross-sections of $e^+e^-\to \gamma X(3872)$\cite{Yuan:2013lma,Ablikim:2013dyn}
\begin{eqnarray}
\label{Eq:3872Bes3}
 \sigma[e^+e^- \to \gamma X(3872)]\times  {\rm Br}[J/\psi \pi\pi]<0.13
  {\rm pb\ \ at \ 90\% \ CL.\ } &&\hspace{0.3cm} \sqrt s= 4.009 {\rm GeV}\nonumber \\
 \sigma[e^+e^-\to \gamma X(3872)]\times  {\rm Br}[  J/\psi \pi\pi]= 0.32
  \pm 0.15 \pm 0.02 {\rm pb} &&\hspace{0.3cm} \sqrt s= 4.230 {\rm GeV}\nonumber \\
 \sigma[e^+e^-\to \gamma X(3872)]\times  {\rm Br}[ J/\psi \pi\pi]=0.35
  \pm 0.12 \pm 0.02{\rm pb} &&\hspace{0.3cm} \sqrt s= 4.260 {\rm GeV}\nonumber \\
 \sigma[e^+e^-\to \gamma X(3872)]\times  {\rm Br}[ J/\psi \pi\pi]<0.39
  {\rm pb\ \ at \ 90\% \ CL.\ } &&\hspace{0.3cm} \sqrt s= 4.360 {\rm GeV}
\end{eqnarray}
Where $ {\rm Br}[  J/\psi \pi\pi]$ means $  {\rm Br}[ X(3872)\to J/\psi \pi\pi]$.
And the studies of $\psi(4160) \to X(3872) \gamma$ \cite{Margaryan:2013tta} and
$\psi(4260) \to X(3872) \gamma$ \cite{Guo:2013zbw} are proposed to
 probe the molecular content of the $X(3872)$.

Many NLO relativistic and radiative corrections for heavy
quarkonium production are considered within nonrelativistic QCD
(NRQCD)\cite{Bodwin:1994jh}.
By introducing the color octet mechanism, one can obtain the infrared-safe calculations
for the decay rates of P wave
\cite{Brambilla:2008zg,Lansberg:2009xh,Hwang:2010iq} and
D wave\cite{He:2008xb,He:2009bf,Fan:2009cj}  quarkonium states.
The color octet contributions of the diphoton decay of P wave quarkonium states
are calculated in Ref.\cite{Ma:2002ev}.
$O(\alpha_{s}v^2)$ corrections to
the decays of $h_c, h_b$ and $\eta_b$ are studied in Ref.\cite{Guo:2011tz,Li:2012rn}.
The NLO QCD corrections\cite{Zhang:2006ay,Wang:2011qg,Zhang:2008gp,Zhang:2009ym,
Gong:2007db,Gong:2008ce,Gong:2009ng,Gong:2009kp,Ma:2008gq,Dong:2011fb,
Bodwin:2013ys},
relativistic corrections\cite{He:2007te,Bodwin:2006ke,Bodwin:2007ga,Elekina:2009wt,Jia:2009np,He:2009uf,Fan:2012dy,Fan:2012vw},
and ${\mathcal O}(\alpha_s v^2)$ corrections \cite{Dong:2012xx,Li:2013qp}
largely compensate for the discrepancies
between theoretical values and experimental measurements at B factories.
The contributions of higher-order QCD corrections for charmonium production
\cite{Campbell:2007ws,Gong:2008hk,Gong:2008sn,Gang:2012js,Ma:2010yw,
Ma:2010vd,Shao:2012iz,Butenschoen:2010rq,Butenschoen:2013pxa,Meng:2013gga}
and polarization \cite{Chao:2012iv,Butenschoen:2012px,Gong:2012ug,Shao:2012fs}
in hadron colliders are also significant.
The relativistic corrections to
$J/\psi$ hadroproduction are significant\cite{Fan:2009zq,Xu:2012am,Li:2013csa}.

We calculate the production of $C=+$ charmonium  at $e^+e^-$ annihilation  at  BESIII
to test the nature of $C=+$ $XYZ$ states.
Our paper is organized as follows. The calculation framework is given in Sec.~\ref{sec:frame}.
The numerical results of the cross-sections of $C=+$ charmonium are discussed in Sec.~\ref{sec:NumCpp}.
A discussion of $X(3872)$ and other  $C=+$~ $XYZ$ states  is given in Sec.~\ref{sec:c+xyz}.
The summary is given in Sec.~\ref{sec:summary}.

\section{The frame of the calculation}\label{sec:frame}

In the NRQCD factorization framework, we can express the amplitude in the rest
frame of $H$ as\cite{Chung:2008km,Li:2009ki,
Sang:2009jc}
\begin{eqnarray}
\label{eq:qqlevel}  &&\hspace{-2.2cm} {\cal A}(e^-(k_1) e^+ (k_2)
\rightarrow H_{c\bar{c}}({}^{2S+1}   L  _{J} )(2p_1)
+\gamma) \nonumber
\\
&=&\sum\limits_{L_{ z} S_{ z} }\sum\limits_{s_1s_2}\sum\limits_{jk}\int {\rm d}^3 \vec{q}
\Phi_{c\bar c}(\vec{q})
\langle s_1;s_2\mid S S_{ z}\rangle 
\langle
3j;\bar{3}k\mid 1\rangle\nonumber\\
&&\times{\cal A}\left[e^-(k_1) e^+ (k_2)\rightarrow
 c_j^{s_1}(p_1+q)+\bar{c}^{s_2}_k(p_1-q)+
 \gamma(k)\right],
\end{eqnarray}

where $\langle 3j;\bar{3}k\mid 1\rangle =\delta_{jk}/\sqrt{N_c}$,
$\langle s_1;s_2\mid S S_{ z}\rangle$ is the color   Clebsch-Gordan coefficient for
$c\bar{c}$ pairs projecting out appropriate bound states, and
$\langle s_1;s_2\mid S S_{ z}\rangle$ is the spin  Clebsch-Gordan coefficient.
${\cal A}\left[e^-(k_1) e^+ (k_2)\rightarrow
 c_j^{s_1}(p_1+q)+\bar{c}^{s_2}_k(p_1-q)+
 \gamma(k)\right]$ is the quark level scattering
 amplitude.
In the rest frame of $H$, $q=(0,\vec{q})$, and $p_1=(\sqrt{m_c^2+\vec{q}^2},0,0,0)$.
$\Phi^H_{c\bar c}(\vec{q})$ is the $c \bar c$ component wave function of hadron $H$ in momentum space.
For $v^2=\vec{q}^2/ m_c^2  \ll 1$\cite{Bodwin:1994jh}, we can expand Eq.(\ref{eq:qqlevel}) with $v^2$:
\begin{eqnarray}
{\cal A}(q)&=& {\cal A}(0)+\left.\frac{\partial  {\cal A}(\vec{q})}{\partial \vec{q}^\alpha}
\right|_{q=0} \vec{q}^\alpha+\left.\frac{\partial^2  {\cal A}(\vec{q})}{\partial
\vec{q}^\alpha\partial \vec{q}^\beta}\right|_{q=0} \frac{ \vec{q}^\alpha  \vec{q}^\beta}{2}
 \nonumber \\
&&  +\left.\frac{\partial^3  {\cal A}(\vec{q})}{\partial \vec{q}^\alpha\partial \vec{q}^\beta
\partial \vec{q}^\delta}\right|_{q=0} \frac{\vec{q}^\alpha
 \vec{q}^\beta
 \vec{q}^\delta}{3!}+....
\end{eqnarray}
Here ${\cal A}(q)={\cal A}\left[e^-(k_1) e^+ (k_2)\rightarrow
c_j^{s_1}(p_1+q)+\bar{c}^{s_2}_k(p_1-q)+
\gamma(k)\right]$. We consider the Fourier transform between the momentum space and position space as:
\cite{Bodwin:1994jh,Xu:2012am},
\begin{eqnarray}
\int {\rm d}^3 \vec{q}\ \ \Phi_{c\bar c}(\vec{q}) &\propto&\sqrt{Z_{c\bar c}^H} R_{c\bar c}(0) \nonumber \\
\int {\rm d}^3 \vec{q}\ \ \vec{q}^\alpha \Phi_{c\bar c}(\vec{q}) &\propto&
\sqrt{Z_{c\bar c}^H} R^\prime_{c\bar c}(0)\nonumber \\
\int {\rm d}^3 \vec{q}\ \   \vec{q}^\alpha  \vec{q}^\beta \Phi_{c\bar c}(\vec{q}) &
\propto&\sqrt{Z_{c\bar c}^H} R^{\prime\prime}_{c\bar c}(0)\nonumber \\
\int {\rm d}^3 \vec{q}\ \ \vec{q}^\alpha
 \vec{q}^\beta
 \vec{q}^\delta \Phi_{c\bar c}(\vec{q}) &\propto&\sqrt{Z_{c\bar c}^H} R^{\prime\prime\prime}_{c\bar c}(0).
\end{eqnarray}
Here $Z_{c\bar c}^H$ is the possibility of $c \bar c$ component in hadron $H$.
$R_{c\bar c}(0)$ is the radial Schrodinger wave function at the origin.  $R^{l}_{c\bar c}(0)$ is
the derivative of the radial  Schrodinger wave function at the origin
\begin{eqnarray}
R^{l}_{c\bar c}(0)=\left.\frac{{\rm d}^l R_{c\bar c}(r)}{{\rm d}^l r}\right|_{r=0}
\end{eqnarray}
$ R_{c\bar c}(0)$ corresponds to the ${\cal O}(v^0)$ S wave matrix element,
$ R^{\prime}_{c\bar c}(0)$ corresponds to the ${\cal O}(v^0)$ P wave matrix element,
$ R^{\prime\prime}_{c\bar c}(0)$ corresponds to the ${\cal O}(v^2)$ S wave matrix element
or ${\cal O}(v^0)$ D wave matrix element, and
$ R^{\prime\prime \prime}_{c\bar c}(0)$ corresponds to the ${\cal O}(v^2)$ P wave matrix element.

$ R_{c\bar c}(0)$ is also written as long-distance matrix elements (LDMEs) as discussed in Ref.\cite{Xu:2012am}. For example,
\begin{eqnarray}\label{Eq:paraxc1}
 &&\langle0|\mathcal{O}^{\chi_{c1}}(^3P_1^{[1]})|0\rangle=\frac{27}{2\pi}|R^\prime_{1P}(0)|^2,
\end{eqnarray}
We calculated the relativistic corrections for the S wave and P wave states and obtain two
LDMEs for $\eta_c$, four LDMEs for $\chi_{cJ}$, and one LDMEs for $^1D_2$ states.
To simplify the discussion of the
numerical result, we assumed that
\begin{eqnarray}
<0|\mathcal{O}^{\chi_{cJ}}({}^3P_J^{[1]})|0>&=&
(2J+1)<0|\mathcal{O}^{\chi_{cJ}}({}^3P_0^{[1]})|0> . \end{eqnarray}
\begin{equation}\label{eq:8ME}
v^2=\frac{\langle0|\mathcal{P}^{H}(^{2s+1}L_J^{[c]})|0\rangle}
{m_c^2\langle0|\mathcal{O}^{H}(^{2s+1}L_J^{[c]})|0\rangle}.
\end{equation}
Then there is only one LDME for $S$ wave, $P$ wave, and $D$ wave respectively.
More details can be found in Ref.\cite{Xu:2012am}.

The relativistic correction $K$ factor is
\begin{eqnarray}
K_{v^2}[\eta_c]&=&-\frac{5 v^2}{6} -\frac{r v^2}{1-r} , \nonumber \\
K_{v^2}[\chi_{c0}]&=&-\frac{\left(55 r^2-28
   r+13\right) v^2}{10 \left(3 r^2-4
   r+1\right)} -\frac{r v^2}{1-r}, \nonumber \\
K_{v^2}[\chi_{c1}]&=&-\frac{\left(21 r^2+30
   r-11\right) v^2}{10
   \left(r^2-1\right)} -\frac{r v^2}{1-r},\nonumber \\
K_{v^2}[\chi_{c2}]&=&-\frac{\left(90 r^3+113
   r^2+4 r-7\right) v^2}{10 (r-1) \left(6 r^2+3
   r+1\right)}  -\frac{r v^2}{1-r},
\end{eqnarray}
where $r=4m_c^2/s$.  $ -\frac{r v^2}{1-r}$ is the relativistic correction
of the phase space. If we select $r \to 0$, the $K_{v^2}$ factor
is consistent with the $K$ factor at large $p_T$ in Ref.\cite{Xu:2012am}.

Our leading order (LO) cross-sections of  $e^+e^-\to
\gamma ^*\to \gamma+H$ is consistent  with Ref.\cite{Chung:2008km,Li:2009ki,
Sang:2009jc}. The QCD corrections of $e^+e^-\to
\gamma ^*\to \gamma+H$ is consistent  with Ref.\cite{Li:2009ki,
Sang:2009jc}.
And
the relativistic corrections of $e^+e^-\to
\gamma ^*\to \gamma+\eta_c$  is consistent  with Ref.\cite{Sang:2009jc,Fan:2012dy,Fan:2012vw}.

We can obtain a similar amplitude for the $D \bar D$ component in the molecule model.
We can estimate the off-resonance amplitude of
$e^+ e^- \to H + \gamma$ from the  $D \bar D$ component.
The parton-level amplitudes may be compared with the hadron-level amplitudes:
\begin{eqnarray}
{\cal A}\left[e^-(k_1) e^+ (k_2)\rightarrow
c\bar{c}(2p_1)+
\gamma\right]  \sim  {\cal A}\left[e^-(k_1) e^+ (k_2)\rightarrow
D\bar{D}(2p_1)+
\gamma\right]
\end{eqnarray}
By contrast, the $ R^{l}_{c\bar c}(0) \sim v^{2l}R^S_{c\bar c}(0)  \gg R_{D\bar D}(0)$ with the $S$ wave $l=0$ and $P$ wave $l=1$ for the binding
energies of $c \bar c$ and $D \bar D$
are  several hundreds of MeV  and several MeV, respectively. If $Z_{c\bar c}^H \sim Z_{D\bar D}^H$,
we can consider the $c \bar c$ contributions only.

In the numerical calculation, we consider the charm quark mass as half of the hadron mass consistent with the physics phase space.
With a large charm quark mass, the wave functions at the origin
are identified as the Cornell potential result in Ref.\cite{Eichten:1995ch}.
The sellected parameters are as follows:
\begin{eqnarray}
&&m_c=m_H/2, \hspace{2cm} \alpha_s=0.23, \hspace{2.4cm}\alpha=1/133, \nonumber \\
&&v^2=0.23, \hspace{2.45cm} R_{1S}=1.454 {\rm GeV}^3, \hspace{0.9cm} R_{2S}=0.927 {\rm GeV}^3, \nonumber \\
&&R_{3S}=0.791 {\rm GeV}^3, \hspace{1cm} R'_{1P}=0.131 {\rm GeV}^5, \hspace{0.8cm}
R'_{2P}=0.186 {\rm GeV}^5, \nonumber \\
&& R''_{1D}=0.031 {\rm GeV}^7.
\end{eqnarray}

The wave functions at origin for higher states are estimated as
\begin{eqnarray}
R_{4S}&=&2\times R_{3S} -R_{2S} = 0.655 {\rm GeV}^3, \nonumber \\
 R'_{3P}&=&(R'_{1P}+R'_{2P})/2=0.159 {\rm GeV}^5, \nonumber \\
  R''_{2D}&=& R''_{1D}=0.031 {\rm GeV}^7.
 \end{eqnarray}

In the numerical result,  "$\sigma_{LO}$" is the LO cross-section, "$\sigma_{v^2}$" is the cross-section including  the LO and the relativistic correction, "$\sigma_{\alpha_s}$" is the cross-section including  the LO and  the radiative correction, and "$\sigma_{
\alpha_s, v^2}$" is the cross-section including  the LO, the relativistic correction, and the radiative correction.
In addition, "LO" is the LO cross-section, "RC" is the relativistic correction, "QCD" is the radiative correction, and "Total"  is the cross-section including the LO, the relativistic correction, and the radiative correction.

For the LO, the cross-section is ${\cal O} (\alpha_s^0 v^0)$.  As $\alpha_s=0.23 \pm 0.03$ and $v^2=0.23 \pm 0.03$ are reasonable estimates, we can estimate that the uncertainty of the numerical result from $\alpha_s$ and $v^2$ is $<10\%$.

\section{Pure $C=+$ charmonium states} \label{sec:NumCpp}
We can estimate the cross-sections for pure  $C=+$ charmonium states $H$ in $e^+e^-\to
\gamma~+~H$ at BESIII with $H=\eta_c(nS)$ (n=1, 2, 3, and 4),
$\chi_{cJ}(nP)$ (n=1, 2, and 3), and $^1D_2(nD)$ (n=1 and 2).  The mass of the lower states can be found in Ref.\cite{Beringer:1900zz}, and the mass of the higher states is  selected from Ref.\cite{Li:2009zu}.

\begin{figure}[ht]
\begin{center}
\includegraphics[width=0.8\textwidth]{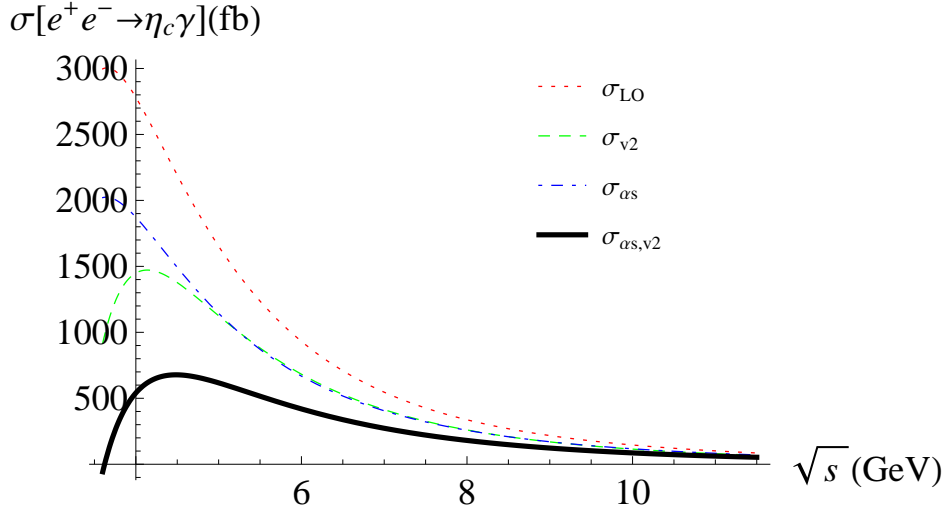}
\end{center}
\caption{\label{fig:etac}The cross-sections of $e^+e^- \to \eta_c + \gamma$ as a function of $\sqrt s$ in fb. The cross-section  "$\sigma_{LO}$", "$\sigma_{v^2}$", "$\sigma_{\alpha_s}$", and "$\sigma_{
\alpha_s, v^2}$" are defined near the end of Section 2.}
\end{figure}

\begin{figure}[ht]
\begin{center}
\includegraphics[width=0.8\textwidth]{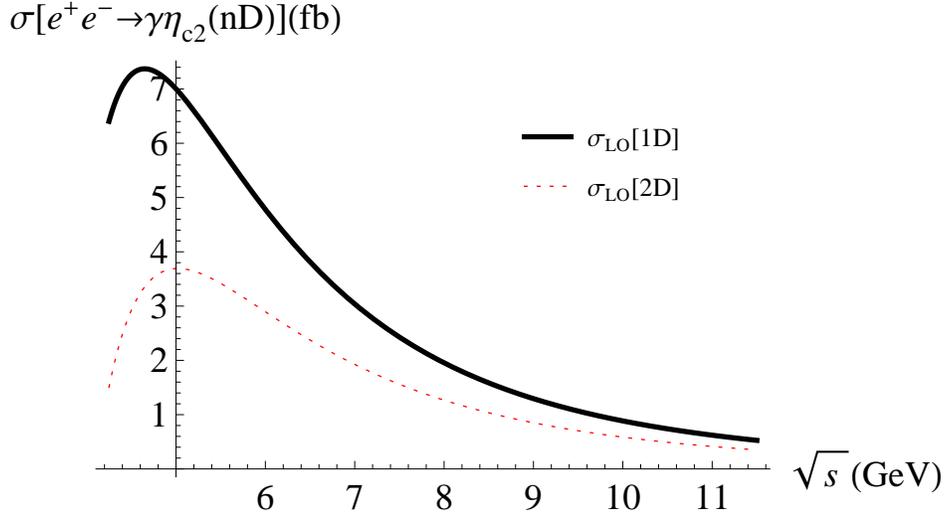}
\end{center}
\caption{\label{fig:etac2D}The cross-sections of $e^+e^-
\to \eta_{c2}(1D,2D) + \gamma$ as a function of $\sqrt s$ in fb.}
\end{figure}

\begin{table*}[htbp]
\caption{The cross-sections of $e^+e^- \to H + \gamma$ for $\eta_c(nS)$
with $n=1,2,3,4$ and $\eta_{c2}{(nD)}$ for $n=1,2$ charmonium states in fb. The labels LO, RC, QCD and Total are defined near the end of Section 2.
The mass of $\eta_{c}(3S)$, $\eta_{c}(4S)$, $\eta_{c2}(1D)$,
and $\eta_{c2}(2D)$ are selected from Ref.\cite{Li:2009zu}.
The other mass can be found in Ref.\cite{Beringer:1900zz}.
\label{tab:etaBESB} }
\centering
\begin{tabular}{cc|ccccccc}
\hline
\multicolumn{2}{c|}{$\sqrt s$(GeV)}
& 4.00 & 4.25  & 4.50  & 4.75  & 5.00  & 10.6 & 11.2 \\ \hline
{$\eta_c$(2981)}    & LO   & 2781 & 2494 & 2192 & 1906 & 1652 & 117 & 95  \\
                    & RC   & -1332& -1033& -814 & -650 & -526 & -25 & -20 \\
                    & QCD  & -909 & -807 & -700 & -598 & -508 & -22 & -16 \\
                    & Total&  540 &  653 & 678  & 658  & 617  &  70 & 58  \\  \hline
{$\eta_c(2S)$(3639)}    &LO& 563  &  684 & 706  & 679  & 629  &  58 & 48  \\
                       & RC& -730 & -563 & -442 & -352 & -284 & -13 & -10 \\
                      & QCD& -177 & -221 & -231 & -222 & -205 & -13 & -10 \\
                    & Total& -344 & -100 &  33  & 105  & 141  &  32 & 27  \\  \hline
{$\eta_c(3S)$(3994)}    &LO&      & 233  & 337  & 374  & 377  &  44 & 36  \\
                       & RC&      & -450 & -352 & -279 & -225 &  -10& -8  \\
                      & QCD&      & -72  & -107 & -121 & -123 &  -10& -8  \\
                    & Total&      & -228 & -122 & -27  &  29  &  24 & 20  \\  \hline
{$\eta_c(4S)$(4250) }   &LO&      &      & 133  & 198  & 225  &  34 & 28  \\
                       & RC&      &      & -279 & -221 & -178 &  -8 & -6  \\
                      & QCD&      &      & -41  & -63  & -73  &  -8 & -7  \\
                    & Total&      &      & -186 & -86  & -26  &  17 & 15  \\  \hline
{$\eta_{c2}(1D)$(3796)} &LO& 4.0  & 6.4  & 7.3  & 7.3  & 7.0  & 0.71& 0.58\\ \hline
{$\eta_{c2}(2D)$(4099)} &LO&      & 1.5   & 2.9 & 3.5  & 3.7  & 0.47& 0.38\\ \hline
\end{tabular}
\end{table*}

The cross-section of $e^+e^- \to \eta_c + \gamma$ as a function of $\sqrt s$
is shown in Fig.\ref{fig:etac}. The cross-sections of $e^+e^-
\to \eta_{c2}(1D,2D) + \gamma$ as a function of $\sqrt s$ are shown in Fig.\ref{fig:etac2D}. The numerical results for $nS$ with $n=1, 2, 3, 4$ and $nD$ with $n=1, 2$
are listed in Table \ref{tab:etaBESB}.  We  determined that the radiative and relativistic corrections are negative and large for $\eta_c(nS)$, respectively. The LO cross-sections for $\eta_{c2}(1D,2D)$ is very small at BESIII; hence, the high order corrections are ignored.

\begin{figure}[ht]
\begin{center}
\includegraphics[width=0.8\textwidth]{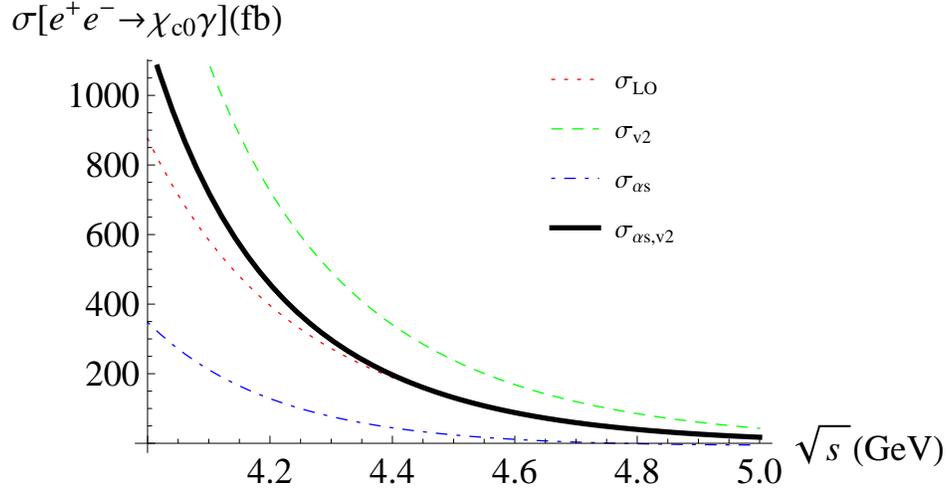}
\end{center}
\caption{\label{fig:chic05}The cross-sections of $e^+e^- \to \chi_{c0} + \gamma$ as a function of $\sqrt s$ in fb. The cross-section  "$\sigma_{LO}$", "$\sigma_{v^2}$", "$\sigma_{\alpha_s}$", and "$\sigma_{
\alpha_s, v^2}$" are defined near the end of Section 2.}
\end{figure}

\begin{figure}[ht]
\begin{center}
\includegraphics[width=0.8\textwidth]{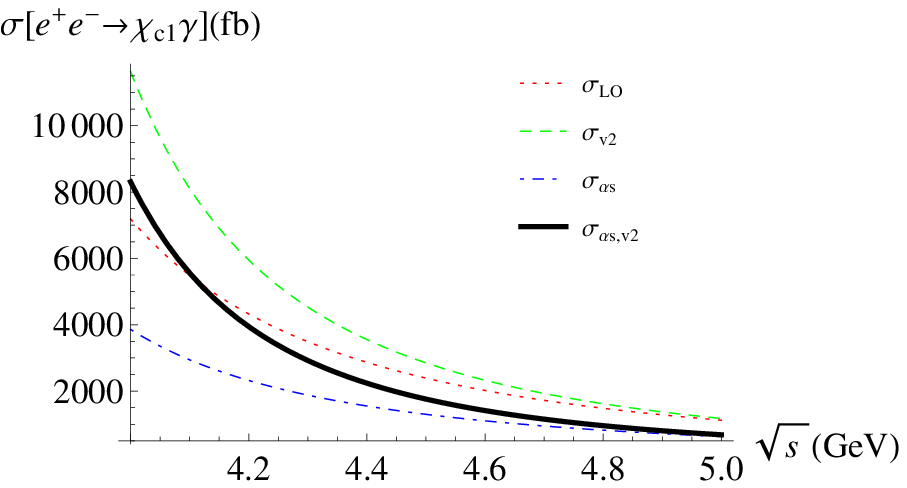}
\end{center}
\caption{\label{fig:chic15}The cross-sections of $e^+e^- \to \chi_{c1}
+ \gamma$ as a function of $\sqrt s$ in fb. The cross-section  "$\sigma_{LO}$", "$\sigma_{v^2}$", "$\sigma_{\alpha_s}$", and "$\sigma_{
\alpha_s, v^2}$" are defined near the end of Section 2.}
\end{figure}

\begin{figure}[ht]
\begin{center}
\includegraphics[width=0.8\textwidth]{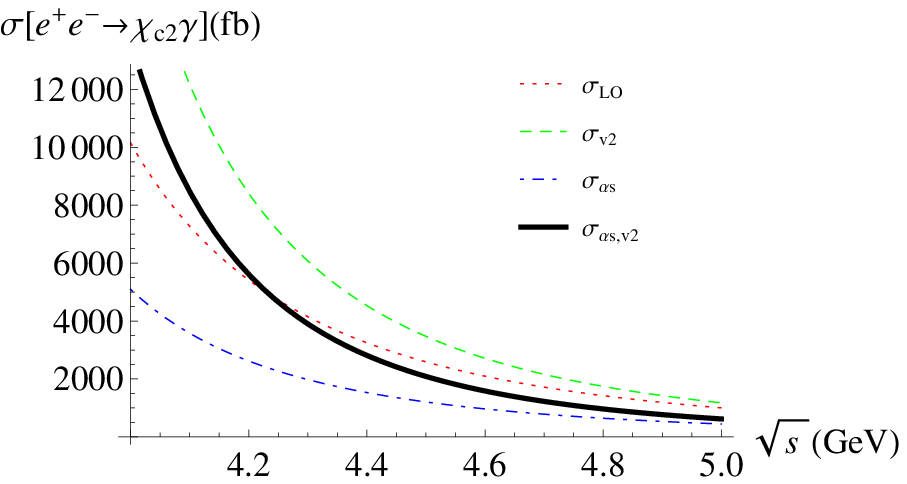}
\end{center}
\caption{\label{fig:chic25}The cross-sections of $e^+e^- \to \chi_{c2} +
\gamma$ as a function of $\sqrt s$ in fb. The cross-section  "$\sigma_{LO}$", "$\sigma_{v^2}$", "$\sigma_{\alpha_s}$", and "$\sigma_{
\alpha_s, v^2}$" are defined near the end of Section 2.}
\end{figure}

The cross-sections of $e^+e^- \to \chi_{cJ} + \gamma$ as a
function of $\sqrt s$ are shown in Fig.\ref{fig:chic05}, Fig.\ref{fig:chic15}, and Fig.\ref{fig:chic25} for $J=0,1,2$,  respectively.
The numerical results for $\chi_{cJ}(nP)$ with $n=1,2,3$
are listed in Table \ref{tab:chic0BESB}, Table \ref{tab:chic1BESB}, and Table \ref{tab:chic2BESB} for $J=0,1,2$, respectively.
We determined that the QCD corrections are large but negative and the relativistic corrections are
large and positive. Hence, many $P$ wave  states can be searched at BESIII.

\begin{table*}[htbp]
\caption{The cross-sections of $e^+e^- \to \chi_{c0}(nP) + \gamma $ with $n=1,2,3$  in fb. The labels LO, RC, QCD and Total are defined near the end of Section 2.
The $\chi_{c0}(2P)$ is considreed as  $X(3915)$($X(3945)$/$Y(3940)$)
\cite{Eidelman:2012vu,Brambilla:2010cs}.
The mass of $\chi_{c0}(3P)$ are selected from Ref.\cite{Li:2009zu}.
The other mass can be found in Ref.\cite{Beringer:1900zz}.
\label{tab:chic0BESB} }
\centering
\begin{tabular}{cc|ccccccc}
\hline
\multicolumn{2}{c|}{$\sqrt s$(GeV)}
& 4.00 & 4.25  & 4.50  & 4.75  & 5.00  & 10.6 & 11.2 \\ \hline
{$\chi_{c0}$(3415)} & LO   &  877 &  328 &  132 & 53   & 21   & 1.81& 1.6 \\
                    & RC   &  825 &  268 &  107 &  48  & 22   &-0.77&-0.63\\
                    & QCD  & -528 & -228 & -107 & -52  & -26  &-0.38&-0.29\\
                    & Total& 1173 &  368 & 131  &  49  &  17  & 1.42&1.22 \\  \hline
{$\chi_{c0}(2P)$(3918)} &LO&      & 1991 & 665  & 271  & 119  & 1.30&1.18 \\
                       & RC&      & 3102 & 680  & 230  & 96   &-0.64&-0.54\\
                      & QCD&      &-1013 & -384 & -177 & -89  & 0.39& 0.30\\
                    & Total&      & 4080 & 962  & 324  & 127  & 1.04& 0.94\\  \hline
{$\chi_{c0}(3P)$(4131)} &LO&      &      & 1073 & 384  & 164  & 0.82& 0.75\\
                       & RC&      &      & 1600 & 391  & 140  &-0.44&-0.38\\
                      & QCD&      &      & -551 & -223 & -107 & 0.29&0.23 \\
                    & Total&      &      & 2121 & 554  & 198  & 0.67&0.61 \\  \hline
\end{tabular}
\end{table*}

\begin{table*}[htbp]
\caption{The cross-sections of $e^+e^- \to \chi_{c1}(nP) + \gamma$ with $n=1,2,3$  in fb. The labels LO, RC, QCD and Total are defined near the end of Section 2.
The mass of $\chi_{c1}(2P,3P)$ are selected from Ref.\cite{Li:2009zu}.
And the mass of $\chi_{c1}(1P)$  can be found in Ref.\cite{Beringer:1900zz}.
\label{tab:chic1BESB} }
\centering
\begin{tabular}{cc|ccccccc}
\hline
\multicolumn{2}{c|}{$\sqrt s$(GeV)}
& 4.00 & 4.25  & 4.50  & 4.75  & 5.00  & 10.6 & 11.2 \\ \hline
{$\chi_{c1}$(3511)} & LO   & 7186 & 3874 & 2392 & 1597 &1124  & 23.5&18.5 \\
                    & RC   & 4448 & 1296 &  459 & 168  & 52   &-4.8 &-3.8 \\
                    & QCD  &-3327 &-1791 &-1091 &-715  & -492 &-6.5 &-4.9 \\
                    & Total& 8307 & 3379 & 1760 & 1051 & 685  & 12.3&9.7  \\  \hline
{$\chi_{c1}(2P)$(3901)} &LO&      & 8854 & 4244 & 2495 & 1624 & 25.7&20.0 \\
                       & RC&      & 9585 & 2297 & 789  & 312  &-4.9 &-3.9 \\
                      & QCD&      & -4041& -1967& -1152& -741 &-7.7 &-5.70\\
                    & Total&      &14397 & 4573 & 2131 & 1195 &13.2 & 10.3\\  \hline
{$\chi_{c1}(3P)$(4178)} &LO&      &      & 1073 & 384  & 164  & 0.82& 0.75\\
                       & RC&      &      & 1600 & 391  & 140  &-0.44&-0.38\\
                      & QCD&      &      & -551 & -223 & -107 & 0.29&0.23 \\
                    & Total&      &      & 2121 & 554  & 198  & 0.67&0.61 \\  \hline
\end{tabular}
\end{table*}

\begin{table*}[htbp]
\caption{The cross-sections of $e^+e^- \to \chi_{c2}(nP) + \gamma$ with $n=1,2,3$  in fb. The labels LO, RC, QCD and Total are defined near the end of Section 2.
$\chi_{c2}(2P)$ is considreed as $Z(3930)$, \cite{Eidelman:2012vu,Brambilla:2010cs}.
The mass of $\chi_{c2}(3P)$ are selected from Ref.\cite{Li:2009zu}. And the mass of
$\chi_{c2}(1P)$  can be found in Ref.\cite{Beringer:1900zz}.
\label{tab:chic2BESB} }
\centering
\begin{tabular}{cc|ccccccc}
\hline
\multicolumn{2}{c|}{$\sqrt s$(GeV)}
& 4.00 & 4.25  & 4.50  & 4.75  & 5.00  & 10.6 & 11.2 \\ \hline
{$\chi_{c2}$(3556)} & LO   &10149 & 4724 & 2590 & 1562 & 1004 & 9.66&7.37 \\
                    & RC   & 8587 & 2385 &  880 & 376  & 173  &-1.16&-0.93\\
                    & QCD  &-5056 &-2455 &-1384 &-851  & -557 &-6.27&-4.82\\
                    & Total&13679 & 4655 & 2087 & 1086 & 621  & 2.22&1.63 \\  \hline
{$\chi_{c2}(2P)$(3927)} &LO&      &13419 & 5581 & 2931 & 1927 &11.29&8.53 \\
                       & RC&      &17835 & 3965 & 1355 & 565  &-1.22&-0.99\\
                      & QCD&      &-6423 & -2822& -1533& -926 &-7.25&-5.52\\
                    & Total&      &24862 & 6723 & 2754 & 1368 &2.82 & 2.03\\  \hline
{$\chi_{c2}(3P)$(4208)} &LO&      &      & 8938 & 3607 & 1886 & 8.55& 6.40\\
                       & RC&      &      &14212 & 2949 & 995  &-0.83&-0.68\\
                      & QCD&      &      & -4210& -1803& -977 &-5.43&-4.10\\
                    & Total&      &      & 18941& 4753 & 1904 & 2.28&1.62 \\  \hline
\end{tabular}
\end{table*}

 The NRQCD requires that the energy of photon at the  center of the mass frame of $e^+e^-$
\begin{eqnarray}
E_\gamma= \frac{s-M_H^2}{2 \sqrt s}\sim \sqrt s -M_H +{\cal O}\left[(1 -M_H/\sqrt s)^2\right]
\end{eqnarray}
 be larger than $\Lambda_{QCD}\sim 300\ {\rm MeV}\sim m_c v^2$.
Although this process is a  QED process, the prediction is not reliable and only a reference value
if this requirement is not satisfied.
If we replace photon with gluon,
the soft photon contributions correspond to the long-distance color octet contributions\cite{Bodwin:1994jh,Sang:2009jc}.

\section{$C=+$ $XYZ$ states}
\label{sec:c+xyz}

$X(4160)$ and $Y(4274)$ are found in the B decay $B\to K+ H \to K +\phi J/\psi$ by CDF collaboration\cite{Aaltonen:2011at}. No signal of $X(4160)$ or $Y(4274)$ is reported
by B factories. Hence, the cross-sections for $X(4160)$ or $Y(4274)$ at BESIII may be too small.
The cross-sections of $e^+e^- \to \gamma H$ for $X(3872)$, $X(3940)$,   $X(4160)$, and $X(4350)$ are discussed here. The $1^{--}$ resonance contributions are ignored here.

\subsection{$X(3872)$}

In the light of the mixture state of the $\chi_{c1}(2P)$ and
$D^0\bar{D}^{\star0}$ molecule, the cross-sections of $X(3872)$ at hadron
collides can be expressed as\cite{Meng:2013gga}:
\begin{equation}
 d\sigma[X(3872) \to J/\psi \pi^+\pi^-]=d\sigma[\chi_{c1}(2P)]{\times }k,
\end{equation}
where $k=Z^{X(3875)}_{c\bar{c}}{\times }Br[X(3872) \to J/\psi \pi^+\pi^-]$.
$Br[X(3872) \to J/\psi \pi^+\pi^-]$ is the branching fraction for $X(3872)$
decay to $J/\psi \pi^+\pi^-$. $Z^{X(3875)}_{c\bar{c}}$ is the possibility of the
$\chi_{c1}(2P)$ component in $X(3872)$. And $k=0.018 \pm 0.04 $ \cite{Meng:2005er,Meng:2013gga}.

\begin{figure}[ht]
\begin{center}
\includegraphics[width=0.8\textwidth]{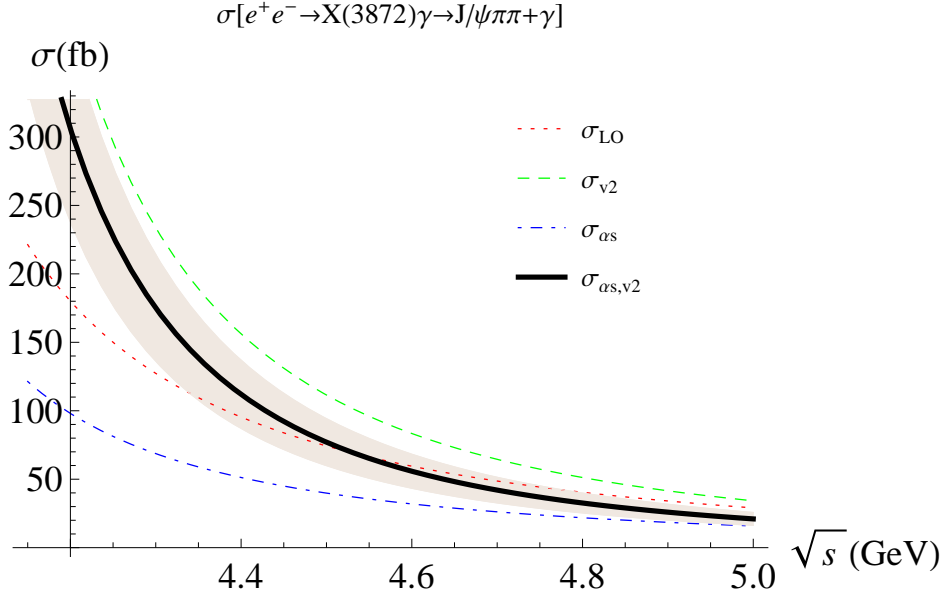}
\end{center}
\caption{\label{fig:x3872}The cross-sections of $e^+e^- \to \chi_{c2} + \gamma$ as a function of $\sqrt s$ in fb. The cross-section  "$\sigma_{LO}$", "$\sigma_{v^2}$", "$\sigma_{\alpha_s}$", and "$\sigma_{
\alpha_s, v^2}$" are defined near the end of Section 2. The uncertainty bind of $\sigma_{
\alpha_s, v^2}$ is from the uncertainty of $k=0.018 \pm 0.04 $.}
\end{figure}

\begin{table*}[htbp]
\caption{The cross-sections of $e^+e^- \to X(3872) + \gamma\to J/\psi \pi\pi + \gamma $  in fb. The labels LO, RC, QCD and Total are defined near the end of Section 2.
\label{tab:x3872} }
\centering
\begin{tabular}{c|ccccccc}\hline
$\sqrt s$(GeV)& 4.15 & 4.2 & 4.25 & 4.3 & 4.35 & 4.45 & 4.55  \\
\hline
LO&  {221$\pm $49} &  {180$\pm $40} &
    {150$\pm $33} &  {127$\pm $28} &
    {110$\pm $24} &    {84$\pm $19} &
    {66$\pm $15}  \\
RC&  {310$\pm $69} &  {208$\pm $46} &
    {146$\pm $32} &  {106$\pm $24} &
    {80$\pm $18} &   {47$\pm $10} &
    {30$\pm $7}  \\
QCD&  {-100$\pm $22} &  {-82$\pm $18} &
    {-69$\pm $15} &  {-59$\pm $13} &
    {-51$\pm $11} &   {-39$\pm $9} &
    {-31$\pm $7}  \\
Total&  {431$\pm $96} &  {306$\pm $68} &
    {227$\pm $51} &  {175$\pm $39} &
    {138$\pm $31} &  {92$\pm $20} &    {65$\pm $14} \\
    \hline
\end{tabular}
\\
\begin{tabular}{c|cc}\hline
$\sqrt s$(GeV)  &  NRQCD prediction for continue  & BESIII \cite{Yuan:2013lma,Ablikim:2013dyn} \\ \hline
4.009           &                   & $<$130  at 90\% CL. \\
4.160           &   $401\pm 89$     &                   \\
4.230           &   $255\pm 57$     & $320 \pm 150 \pm 20$ \\
4.260           &   $215\pm 48$     & $350 \pm 120 \pm 20$\\
4.360           &   $133\pm 29$     & $<$130  at 90\% CL. \\
4.415           &   $105\pm 23$     &         \\
4.660           &   $47\pm 10$      &         \\
    \hline
\end{tabular}

\end{table*}

To clarify the nature of $X(3872)$, we also give the numerical calculation of
$e^+e^- \to \gamma X(3872)\to J/\psi \pi^+\pi^- \gamma$ in this picture
\begin{eqnarray}
&&\sigma[e^+e^-\to \gamma X(3872)]\times  {\rm Br}[ X\to J/\psi \pi\pi]\nonumber \\
&=&
\sigma[e^+e^-\to \gamma \chi_{c1}(2P)(3872)]{\times }(0.018\pm 0.004)
\end{eqnarray}
The cross-sections as a function
of $\sqrt s$ is shown in Fig.\ref{fig:x3872}.
Many $1^{--}$ states with $M_H < 5~$ GeV are also observed. We can predict the cross-sections from continuous contributions at this point, and the result is listed in Table \ref{tab:x3872}. We ignore the $1^{--}$ resonances contributions here.
We emphasize that if we select $\sqrt s= 4.009 {\rm GeV}$, the energy of photon
$ E_\gamma= 134 ~ $ MeV and smaller than $\Lambda_{QCD}\sim m_c v^2\sim 300\ {\rm MeV}$.
 Hence, NRQCD cannot accurately predict the cross-sections with a soft photon with $\sqrt s= 4.009 {\rm GeV}$\cite{Bodwin:1994jh}.
If $\sqrt s= 4.160 {\rm GeV}$, the energy of photon is $E_\gamma=270 {\rm MeV}$.
Although this process is a QED process, the prediction is not reliable and only a reference value\cite{Sang:2009jc}.
We determined that the NRQCD prediction of the continuous contributions can be compared with the BESIII data  of the cross-sections of $e^+e^-\to \gamma X(3872)$
\cite{Yuan:2013lma,Ablikim:2013dyn}  in Eq.(\ref{Eq:3872Bes3}).

When we only considered the continuum production, the resonance contributions can be estimated as that:
\begin{eqnarray}
\sigma_{Res}[s]=\frac{12\pi\Gamma[Res\to e^+e^-]\Gamma[Res \to \gamma X]}{(s-M^2)^2+(M \Gamma_{tot}[Res])^2}.
\end{eqnarray}
We take into account only one resonance here and ignore continuum and other resonances here. If we ignore the interference between one resonance and continuum and other resonances, the $gamma$ energy dependence of the $\Gamma[Res \to \gamma X]$, and $D\bar D$ contributions of decay of $Res \to \gamma X$, we can estimate the resonance contributions. With $X(3872)$ considered as $2P$ states, the largest decay widths are $\psi(4040)$ and $\psi(4160)$, which are considered as the mixing of $\psi(3S)$ and  $\psi(2D)$ \cite{Li:2012vc,Barnes:2005pb}. The $\Gamma[Res \to \gamma X]$ for other states will be less than $1$~keV \cite{Barnes:2005pb}, and $\Gamma_{tot} \sim 100 ~$MeV, $\Gamma[Res\to e^+e^-]\sim 1~$keV. Hence,  we ignore the contributions from other resonances. With the parameters for  $\psi(4040)$ and $\psi(4160)$\cite{Beringer:1900zz,Barnes:2005pb}:
\begin{eqnarray}
&&\Gamma[\psi(4040)\to e^+e^-]=0.87~{\rm keV} \hspace{0.45cm}
   \Gamma[\psi(4040) \to \gamma X]=40~{\rm keV}   \hspace{0.54cm}
     \Gamma_{tot}[\psi(4040)]=80~{\rm MeV} \nonumber \\
&&\Gamma[\psi(4160)\to e^+e^-]=0.83~{\rm keV} \hspace{0.45cm}
   \Gamma[\psi(4160) \to \gamma X]=140~{\rm keV}   \hspace{0.25cm}
     \Gamma_{tot}[\psi(4160)]=103~{\rm MeV} \nonumber
\end{eqnarray}
Hence,  we can determine the contributions of these parameters to $X(3872)\gamma \to J/\psi \pi^+ \pi^- \gamma$
\begin{eqnarray}
&&(\sigma_{\psi(4040)}[4.23]+\sigma_{\psi(4160)}[4.23])\times k= (62 \pm 14) fb\nonumber \\
&&(\sigma_{\psi(4040)}[4.26]+\sigma_{\psi(4160)}[4.26])\times k= (37 \pm 8) fb
\end{eqnarray}
If we considered the interference, the result would be more complex. On the other hand, we have calculated the quark-level intermediate states, which do not clearly deal with the hadron-level intermediate states.

\subsection{$X(3940)$ and  $X(4160)$}\label{sec:x39404160}

\begin{figure}[ht]
\begin{center}
\includegraphics[width=0.8\textwidth]{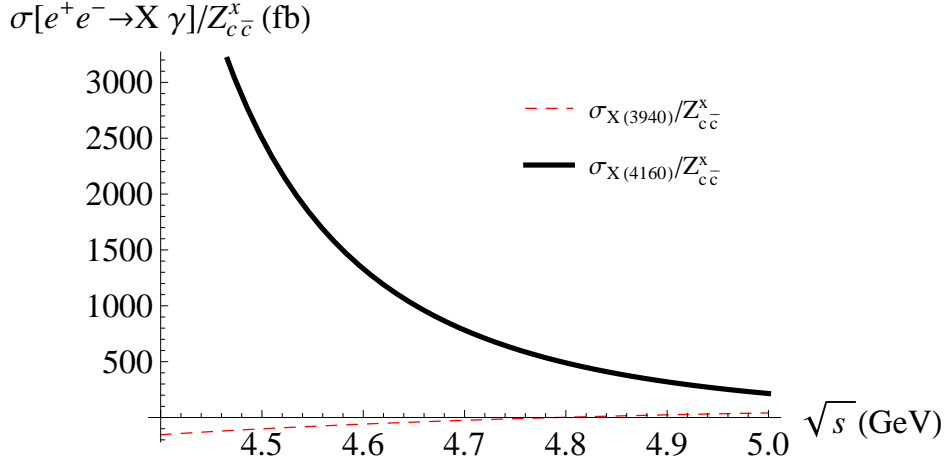}
\end{center}
\caption{\label{fig:x39404160}The cross-sections of $e^+e^- \to X(3940)(X(4160)) + \gamma$ as a function of $\sqrt s$ in fb.}
\end{figure}

$X(3940)$ and  $X(4160)$ are observed in $e^+e^-\to J/\psi\,(  D   \bar D  )$  at B factories \cite{Abe:2007sya}. $\eta_c$ and $\chi_{c0}$ are recoiled with $J/\psi$, but $\chi_{c1}$
and $\chi_{c2}$ are missed\cite{Abe:2007sya}. The theoretical predictions are consistent with the experimental data\cite{Liu:2002wq,Liu:2004ga,Wang:2011qg,Dong:2011fb}.
So there should be large  $\eta_c(nS)$ and
$\chi_{c0}(nP)$  component in $X(3940)$ and $X(4160)$, respectively. The mass of $\eta_c(3S)$ and $\chi_{c0}(3P)$ are predicted as $3994$~MeV and $4130$~MeV respectively\cite{Li:2009zu}.  Compared with Table \ref{tab:etaBESB} and Table \ref{tab:chic0BESB}, we can found that the cross-sections of $\eta_c(3S)$ is small even negative at $\sqrt s< $~5 GeV. But  $\chi_{c0}(3P)$ is large. The cross-sections as a function of $\sqrt s$ is shown in Fig \ref{fig:x39404160}. Here $Z_{c\bar c}^X\le 1$ is the possibility of $\eta_c(3S)$ and
$\chi_{c0}(3P)$  component in $X(3940)$ and  $X(4160)$ respectively. The BESIII collaboration can search $X(3940)$ and $X(4160)$ in the process $e^+e^-\to \gamma\ +X(  D   \bar D  )$. The result may be useful in identifying the nature of   $X(3940)$ and $X(4160)$.

\subsection{$X(4350)$}\label{sec:x4350}

\begin{figure}[ht]
\begin{center}
\includegraphics[width=0.8\textwidth]{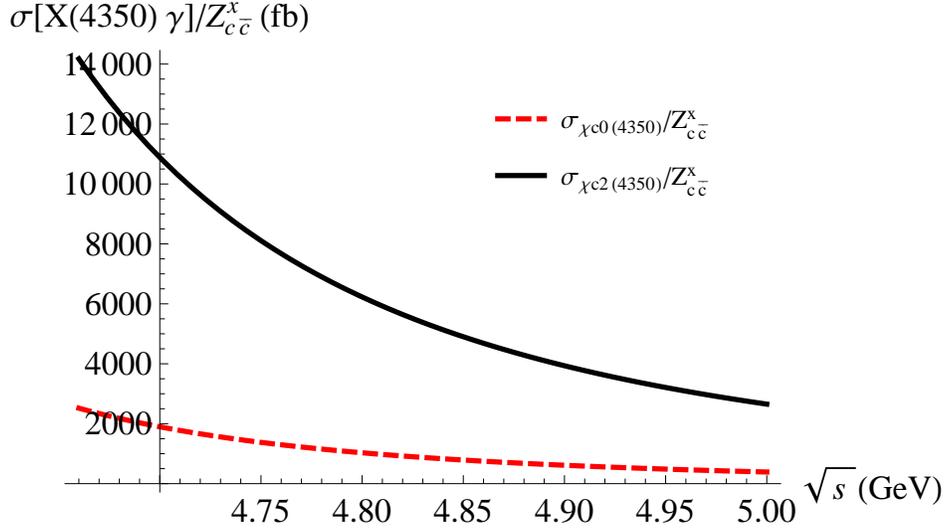}
\end{center}
\caption{\label{fig:chic024350gamma}The cross-sections of $e^+e^- \to X(4350) + \gamma$ as a function of $\sqrt s$ in fb. The cross-section  "$\sigma_{LO}$", "$\sigma_{v^2}$", "$\sigma_{\alpha_s}$", and "$\sigma_{
\alpha_s, v^2}$" are defined near the end of Section 2. And $Z_{c\bar c}^X$ is the possibility of $\chi_{c0}(nP)$ or
$\chi_{c2}(nP)$  component in $X(4350)$.}
\end{figure}

$X(4350)$ are found in $\gamma\gamma \to H \to \phi   J/\psi   $ at B factories \cite{Shen:2009vs}. And $J^{PC}$ is $0^{++}$ or $2^{++}$. So there should be large  $\chi_{c0}(nP)$ or
$\chi_{c2}(nP)$  component in $X(4350)$. In Ref.\cite{Li:2009zu},
The mass of $\chi_{c2}(3P)$ is 4208 MeV. Ignore more detail of the mass, we considered it as $\chi_{c0}(nS)$ or
$\chi_{c2}(nP)$, the wave function at origin  are estimated as
\begin{eqnarray}
R'= R'_{3P}&=&(R'_{1P}+R'_{2P})/2=0.159 {\rm GeV}^5,
\end{eqnarray}
The cross-sections of $e^+e^- \to X(4350) + \gamma$ as a function of $\sqrt s$ is show in Fig.\ref{fig:chic024350gamma}. Here $Z_{c\bar c}^X$ is the possibility of $\chi_{c0}(nP)$ or
$\chi_{c2}(nP)$  component in $X(4350)$. The cross-section for $\chi_{c2}(nP)$ is larger than $\chi_{c0}(nP)$ by a factor of $6$. The result may be useful in  identifying the nature of   $X(4350)$.

\section{Summary and discussion}~\label{sec:summary}
While BESIII and Belle have collected a large amount of data, some final states may be searched by the experimentalists. We can estimate the possible event number at BESIII and Belle.  The possible event number is
\begin{eqnarray}
N=\sigma[e^+e^-\to \gamma + c\bar c[n]]\times  Z_{c\bar c}^H \times Br \times {\cal L} \times \epsilon,
\end{eqnarray}
where $\epsilon$ is the efficiency of detectors selected as $20\%$, $Br$ is the branch ratio of $H$ to the decay mode, and ${\cal L}$ is the luminosity.
The result is listed in Table \ref{tab:cEvenXYZ}.

\begin{table*}[htbp]
\caption{The possible event number of $C=+$ charmonium and $XYZ$ states through $e^+e^-\to \gamma +H$ at  BESIII and Belle. The efficiency of detectors are selected as $20\%$. The integrated luminosity is $1.0 fb^{-1} @ 4.23 $  ~GeV,  $1.0 fb^{-1} @ 4.26$  ~GeV,  $0.5 fb^{-1} @ 4.66$~GeV, and $1 ab^{-1} @ 10.6 $~GeV. The decay mode of $nK m\pi$ corresponds to $D \bar D$ decay,   and the branch ratio is estimated as $1\%$.
\label{tab:cEvenXYZ} }
\centering
\begin{tabular}{c|ccc|c|c|c|c}\hline
H& Decay  & $Br$ & $Z_{c\bar c}^H$   & 4.23   &  4.26   &  4.66  & 10.6   \\
\hline
$\eta_c$    & $K\bar K \pi$    & $7.2 \%$ & 1&    9   & 9  &5  &1012      \\
$\chi_{c0}$ & $2\pi^+ 2\pi^-$  & $2.2 \%$ & 1&   2   &  2  &   &6        \\
$\chi_{c1}$ & $\gamma l^+l^-(\gamma J/\psi)$ & $4.1 \%$ & 1&   29   &  27  &5  &101\\
$\chi_{c2}$ & $\gamma l^+l^-(\gamma J/\psi)$ & $2.3 \%$ & 1&   23   &  20  &3  &10\\
{$\eta_{c2}(1D)$} &$\gamma\gamma K\bar K \pi$ & $1.5 \%$ & 1 &    &  &    & 2 \\
$\eta_c(2S)$ & $K\bar K \pi$ & $1.9\%$ & 1 &        &    &  &  123\\
$X(3872)
(\chi_{c1}(2P))$ & $\pi^+\pi^-l^+l^-(\pi^+\pi^- J/\psi)$  & $0.6 \%$ & 0.36&    6   &  5 &  1 &6        \\
$X(3915)
(\chi_{c0}(2P))$ & $\pi^+\pi^-\pi^0
           l^+l^-(\omega J/\psi)$  & $1 \%$ & 1&    9   & 8 &   &2        \\
$Z(3930)
(\chi_{c2}(2P))$ &  $n K  m \pi(D\bar D)$  & $1 \%$ & 1&   57  & 46 &  4 &6        \\
$X(3940)(\eta_c(3S))$ & $n K  m \pi(D\bar D)$ & $1\%$ & 1 &         &    &  &  48\\
 \hline
\hline

\end{tabular}

\end{table*}

As a summary,
we study the production of $C=+$ charmonium states $H$ in $e^+e^-\to
\gamma~+~H$ at BESIII with $H=\eta_c(nS)$ (n=1, 2, 3, and 4),
$\chi_{cJ}(nP)$ (n=1, 2, and 3), and $^1D_2(nD)$ (n=1 and 2)
within the framework of NRQCD.
The radiative and relativistic corrections are calculated
to next-to-leading order for $S$ and $P$ wave states.  We then argue that the
search for $C=+$ $XYZ$ states such as $X(3872)$, $X(3940)$,  $X(4160)$, and $X(4350)$ in $e^+e^-\to
\gamma~+~H$ at BESIII may help clarify the nature of
these states. BESIII can search $XYZ$ states through two body process $e^+e^-\to \gamma H$, where $H$ decay to $J/\psi \pi^+\pi^-$, $J/\psi \phi$, or $D \bar D$. This result may be useful in identifying the nature of $C=+$ $XYZ$ states.
For completeness, the production of  $C=+$ charmonium
in $e^+e^-\to \gamma +~H$ at B factories is also discussed.

\acknowledgments{
The authors would like to thank Professor C.P. Shen for useful
discussion. This work was
supported by the National Natural Science
Foundation of China (Grants No.11075011 and No. 11375021), the Foundation for the Author of National
Excellent Doctoral Dissertation of China (Grants No. 2007B18 and No. 201020), the Fundamental Research Funds for the Central Universities, and the Education Ministry of
LiaoNing Province.}


\providecommand{\href}[2]{#2}\begingroup\raggedright\endgroup

\end{document}